\documentclass{article_saj}
\usepackage{graphicx,saj,multicol,subeqnarray}
\usepackage{natbib}
\usepackage{float}
\usepackage{xcolor}
\usepackage{widetext}
\usepackage{url}
\usepackage{amsmath,upgreek}
\definecolor{xlinkcolor}{cmyk}{1,0.6,0,0}
\usepackage[bookmarks=false,         
     pdfnewwindow=true,      
     colorlinks=true,    
     linkcolor=xlinkcolor,     
     citecolor=xlinkcolor,     
     filecolor=xlinkcolor,  
     urlcolor=xlinkcolor,      
final=true
]{hyperref}

\def\papertype{\ \hfill\ Original Scientific Paper}

\def\udc{524.7–423}
\renewcommand{\thefootnote}{\fnsymbol{footnote}}

\begin{document}
\parindent=.5cm
\baselineskip=3.1truemm
\columnsep=.5truecm
\newenvironment{lefteqnarray}{\arraycolsep=0pt\begin{eqnarray}}
{\end{eqnarray}\protect\aftergroup\ignorespaces}
\newenvironment{lefteqnarray*}{\arraycolsep=0pt\begin{eqnarray*}}
{\end{eqnarray*}\protect\aftergroup\ignorespaces}
\newenvironment{leftsubeqnarray}{\arraycolsep=0pt\begin{subeqnarray}}
{\end{subeqnarray}\protect\aftergroup\ignorespaces}
%


\markboth{\eightrm DARK MATTER MASS LOSS IN GALAXY FLYBYS}
{\eightrm A. MITRA{\v S}INOVI{\' C}}

\begin{strip}

{\ }

\vskip-1cm

\publ

\type

{\ }


\title{DARK MATTER MASS LOSS IN GALAXY FLYBYS: DEPENDENCE ON IMPACT PARAMETER}


\authors{A Mitra{\v s}inovi{\' c}$^{1,2}$}

\vskip3mm


\address{$^1$Astronomical Observatory, Volgina 7, 11060 Belgrade 38, Serbia}

\address{$^2$Department of Astronomy, Faculty of Mathematics,
University of Belgrade\break Studentski trg 16, 11000 Belgrade,
Serbia}


\Email{amitrasinovic@aob.rs}


\dates{September 15, 2021}{January 26, 2022}


\summary{Galaxy flybys, interactions where two independent halos inter-penetrate but detach at a later time and do not merge, occur frequently at lower redshifts. These interactions can significantly impact the evolution of individual galaxies - from mass loss and shape transformation to the emergence of tidal features and the formation of morphological disc structures. The main focus of this paper is on the dark matter mass loss of the secondary, intruder galaxy, with the goal of determining a functional relationship between impact parameter and dark matter mass loss. Series of N-body simulations of typical galaxy flybys (10:1 mass ratio) with differing impact parameters show that dark matter halo leftover mass of intruder galaxy follows logarithmic growth law with impact parameter, regardless of the way total halo mass is estimated. Lost mass then, clearly, follows exponential decay law. Stellar component stretches faster as impact parameter decreases, following exponential decay law with impact parameter. Functional dependence on impact parameter in all cases seems universal, but fitting parameters are likely sensitive to interaction parameters and initial conditions (e.g. mass ratio of interacting galaxies, the initial relative velocity of intruder galaxy, interaction duration). While typical flybys, investigated here, could not be the sole culprit behind the formation of ultra-diffuse or dark matter deficient galaxies, their effects should not be disregarded as they can at least contribute substantially. Rare, atypical and stronger flybys are worth exploring further.}



\end{strip}

\tenrm


\section{INTRODUCTION}
\indent Galaxy flybys are interactions where two independent halos inter-penetrate but detach at a later time, thus not resulting in a merger. This definition was introduced by \cite{sinha2012}, making a clear distinction between galaxy flybys and close galaxy passages where two halos remain separate at all times. The authors based their analysis on cosmological N-body simulations and found that the number of flybys can even surpass the number of mergers on lower redshifts ($z<\sim 2$). Follow-up study \citep{sinha2015} further explored interaction parameters: in a majority of flybys, secondary halo penetrates deeper than $\sim R_\mathrm{half}$ with initial relative velocity $\sim 1.6 \times V_\mathrm{vir}$ of the primary halo. The typical mass ratio of interacting galaxies was found to be $\sim 0.1$ at high redshifts, or even lower, at the lower redshift end.\\
\indent The frequency and strength of galaxy flybys suggest that these interactions have the potential to significantly impact the evolution of individual galaxies, with the strongest contribution at the present epoch \citep{shan2019}. The focus of previous studies was predominantly on the primary (more massive) galaxy. It was established that flybys can trigger or speed up bar formation \citep{lang2014,lokas2018}, create warps at the edges of the galactic disk, both gaseous and stellar \citep{kim2014} or, in general, reproduce diverse morphology of observed galaxies with differing interaction parameters \citep{pettitt2018}.\\
\indent However, effects on the secondary, intruder galaxy are equally (if not more) important, given the fact that the majority of morphological disturbances seen in dwarf galaxies ($M_\star < 10^9 \mathrm{M}_\odot$) are primarily the result of interactions like these that do not end in a merger \citep{martin2021}. This is particularly noticeable in galaxy clusters. \cite{tormen1998} reported that very close, penetrating encounters between satellites within the cluster are frequent, with almost $60\%$ of satellites experiencing at least one such event before losing $80\%$ of their initial mass. They noted that mass loss, which follows these interactions, is comparable to the one caused by global tides. Thus, interactions within galaxy clusters can contribute to the dynamical evolution of individual galaxies as equally as the global, collective effects of the cluster itself. \cite{gnedin2003} confirmed this, finding that peaks of the tidal force do not always correspond to the closest approach to the cluster centre, but instead to the local density structures (e.g. massive galaxies or the unvirialized remnants of infalling groups of galaxies).\\
\indent Dark matter halos of tidally affected galaxies, due to their extended nature, usually suffer significant mass loss - their outermost parts, being loosely gravitationally bound, get stripped first. With prolonged tidal effects or stronger tidal forces, the inner parts become affected and prone to the tidal stripping. Since tidal forces remove mass from outside in, the process is known as outside-in tidal stripping \citep[e.g.][]{Diemand2007, Choi2009}. At the same time, the stellar counterpart is barely affected \citep{smith2015,smith2016}. Stripping and mass loss of the stellar component typically only starts happening after a significant portion of dark matter ($\sim 80\%$) is already lost \citep{smith2016,lokas2020}. Tidal stripping, in general due to its outside-in nature, is known to be one of the formation mechanisms of ultra-compact dwarfs \citep{bekki2001,bekki2003,pfeffer2013,pfeffer2014,martinovic2017,fm2018,kim2020}, and there is growing evidence for tidal origin of ultra-diffuse galaxies \citep{carleton2019,iodice2021,jones2021,wright2021}. This is all reflected in stellar-to-halo mass relation \citep{niemiec2017,niemiec2019,engler2021} - galaxies in high-density environments, such as clusters, tend to have smaller (than expected) halo masses for a given stellar mass. Exotic class of dark matter deficient galaxies can be considered as an extreme example of these mechanisms \citep{ogiya2018,montes2020,shin2020,jackson2021,maccio2021,trujillogomez2021}.\\
\indent The aim of this paper is to explore the role of impact parameter (which will also be referred to as pericentre distance) in galaxy flybys, with emphasis on dark matter mass loss of the secondary, intruder galaxy. The main goal is to answer a question - is there a functional relationship between impact parameter and dark matter mass loss of the secondary galaxy, and if so, what does it look like? This will be done by utilizing a series of N-body simulations of galaxy flybys with differing impact parameter (described in Section \ref{models}). Total dark matter mass of the intruder galaxy after the encounter will be estimated using three different criteria described in Subsection \ref{mass} As mass loss of the stellar component is not expected, it will be verified if that is the case and if so, changes to its half-mass radius will be explored. Results will be outlined and briefly discussed in Section \ref{results} Finally, Section \ref{conclusion}, besides drawing conclusions, will tackle potential issues of this work, and discuss open questions.

\renewcommand{\thefootnote}{\arabic{footnote}}

\section{MODELS, SIMULATIONS AND METHODS}\label{models}

\indent Two galaxy models were constructed using \texttt{GalactICs} software package \citep{gal95,gal05,gal08}. Primary galaxy model (which will be referred to as galaxy) consists of NFW \citep{nfw1996} dark matter halo, exponential stellar disc and \citet{hern1990} stellar bulge. Dark matter halo, consisting of $N_\mathrm{H} = 6\cdot 10^5$ particles, has total mass $M_\mathrm{H} = 9.057 \cdot 10^{11} \mathrm{M}_\odot$, scale length $a_\mathrm{H}=13.16$ kpc, and concentration parameter $c = 15$. Exponential stellar disc, consisting of $N_\mathrm{D} = 3\cdot 10^5$ particles, has total mass $M_\mathrm{D} = 7.604 \cdot 10^{10} \mathrm{M}_\odot$, scale length $R_\mathrm{D} = 5.98$ kpc, scale height $z_\mathrm{D} = 0.688$ kpc, and central velocity dispersion $\sigma_{R_0} = 98.9$ km s$^{-1}$. Stellar bulge, consisting of $N_\mathrm{B} = 1\cdot 10^5$ particles, has total mass $M_\mathrm{B} = 2.502 \cdot 10^{10} \mathrm{M}_\odot$, and scale radius $R_\mathrm{B} = 2.182$ kpc.\\
\indent Secondary galaxy model (which will be referred to as intruder) consists only of NFW dark matter halo, and stellar bulge, to mimic (dwarf) spherical galaxy for the sake of simplicity. Intruder model was scaled to be $10$ times smaller than galaxy model, in both number of particles and total mass. This results in dark matter halo with $N_\mathrm{H}=6\cdot 10^4$ particles, $M_\mathrm{H} = 9.044 \cdot 10^{10} \mathrm{M}_\odot$ total mass, $a_\mathrm{H} = 4.578$ kpc scale length, and $c = 20$ concentration parameter. Stellar component has, thus, $N_\mathrm{S} = 4\cdot 10^4$ particles, and $M_\mathrm{S} = 1.022 \cdot 10^{10} \mathrm{M}_\odot$ total mass with scale radius $R_\mathrm{B} = 3.145$ kpc. Relevant parameters for both galaxies are listed in Table \ref{tabmod}.\\
\indent Flyby simulations, originally designed to follow the evolution of primary galaxy disc long after the encounter in \citet{mamm2020}, were performed using publicly available code \texttt{GADGET2} \citep{springel2005} compiled with the option to calculate and output particle potential energy. The system was evolved for $5$ Gyr with outputs being saved every $0.01$ Gyr. Galaxy and intruder were initially set as contact system, ie. distance from their centers is roughly equal to the sum of their virial radii $d_0 = R_{\mathrm{vir},1} + R_{\mathrm{vir},2} \approx 290$ kpc, with galaxy being static in the centre of simulation box. Intruder was set on prograde parabolic orbit, co-planar with galaxy disc, with initial relative velocity $v_0 = 500$ km s$^{-1}$. Different pericentre distances (impact parameters - these terms will be used interchangeably) $b$ were achieved in different simulations by slightly varying initial position and velocity angles. In turn, pericentre velocities are also different, while pericentre occurs at the same time ($0.56$ Gyr) and duration of interaction remains the same ($1.08$ Gyr) in all simulations. Interaction is, here, defined as the time during which dark matter halos overlap ($d \leq R_{\mathrm{vir},1} + R_{\mathrm{vir},2}$).\\
\indent Simulation relevant parameters are listed in Table \ref{tab1}. Simulations were named after rough estimate of impact parameters - evidently, they differ from actual impact parameters (column $b$). Based on the results of \citet{sinha2015}, these simulations cover deep flybys, with impact parameter ranging from $0.114 \cdot R_{\mathrm{vir},1}$ to slightly over galaxy's half-mass radius ($\sim 49$ kpc). Mass ratio $q = 0.1$ was found to be the most common one when redshift $z$ is disregarded. However, adopted initial relative velocity is higher than the most common one. \cite{kim2014} used higher initial relative velocity, $v_0 = 600$ km/s, based on the report of \cite{gnedin2003} - in Virgo-type cluster simulation, relative velocities of interacting galaxies show a skewed distribution, peaking at $\sim 350$ km/s, with median $\sim 800$ km/s, and mean value $\sim 1000$ km/s. Thus, value of $v_0 = 500$ km/s can still be considered as representative and realistic. Additionally, attractive perk of this value is that it is high enough to significantly simplify dark matter bound mass estimate (defined in \ref{mass}) making it possible to utilize \texttt{GADGET2}'s calculation of potential energy.\\
\begin{table}
\caption{{List of relevant parameters for two galaxy models where the first block is related to dark matter halo, the second one to stellar bulge and the third to stellar disc.}}\label{tabmod}
\vskip.25cm
\centerline
{\begin{tabular}{ccc}
      & Primary galaxy  & Secondary galaxy \\ \hline \hline
$N_\mathrm{H}$ & $6 \cdot 10^5$ & $6 \cdot 10^4$\\ 
$M_\mathrm{H}$ & $9.057 \cdot 10^{11} \mathrm{M}_\odot$ & $9.044 \cdot 10^{10} \mathrm{M}_\odot$ \\ 
$a_\mathrm{H}$ & $13.16$ kpc & $4.578$ kpc \\ 
$c$    & $15$  & $20$                      \\ \hline 
$N_\mathrm{B}$ & $1\cdot 10^5$  & $4\cdot 10^4$ \\ 
$M_\mathrm{B}$ & $2.502 \cdot 10^{10} \mathrm{M}_\odot$ & $1.022 \cdot 10^{10} \mathrm{M}_\odot$ \\ 
$R_\mathrm{B}$ & $2.182$ kpc & $3.145$ kpc  \\ \hline 
$N_\mathrm{D}$ & $3\cdot 10^5$  &          \\ 
$M_\mathrm{D}$ & $7.604 \cdot 10^{10} \mathrm{M}_\odot$ &                \\ 
$R_\mathrm{D}$ & $5.98$ kpc         &              \\ 
$z_\mathrm{D}$ & $0.688$ kpc        &             \\ 
$\sigma_{R_0}$ & $98.9$ km s$^{-1}$ &        \\ \cline{1-2}
\end{tabular}}
\end{table}
\begin{table}
\caption{{List of flyby simulations where $b$ is pericentre distance, $R_{\mathrm{vir},1}$ virial radius of main galaxy, and $v_b$ pericentre velocity.}}\label{tab1}
\vskip.25cm
\centerline{\begin{tabular}{cccc}
\hline
Name & $b$ [kpc] & $b/R_{\mathrm{vir},1}$ & $v_b$ [km s$^{-1}$]\\
\hline
B30 & 22.50 & 0.114 & 660.14\\
B35 & 26.53 & 0.135 & 650.86\\
B40 & 30.69 & 0.156 & 641.80\\
B45 & 35.07 & 0.178 & 632.86\\
B50 & 39.62 & 0.201 & 624.25\\
B55 & 44.27 & 0.224 & 616.16\\
B60 & 48.99 & 0.248 & 608.09\\
B65 & 53.72 & 0.272 & 601.28\\
\hline
\end{tabular}}
\end{table}
\subsection{Assessment of intruder's center}
\begin{figure}[!ht]
	\centerline{\includegraphics[width=0.95\columnwidth, keepaspectratio]{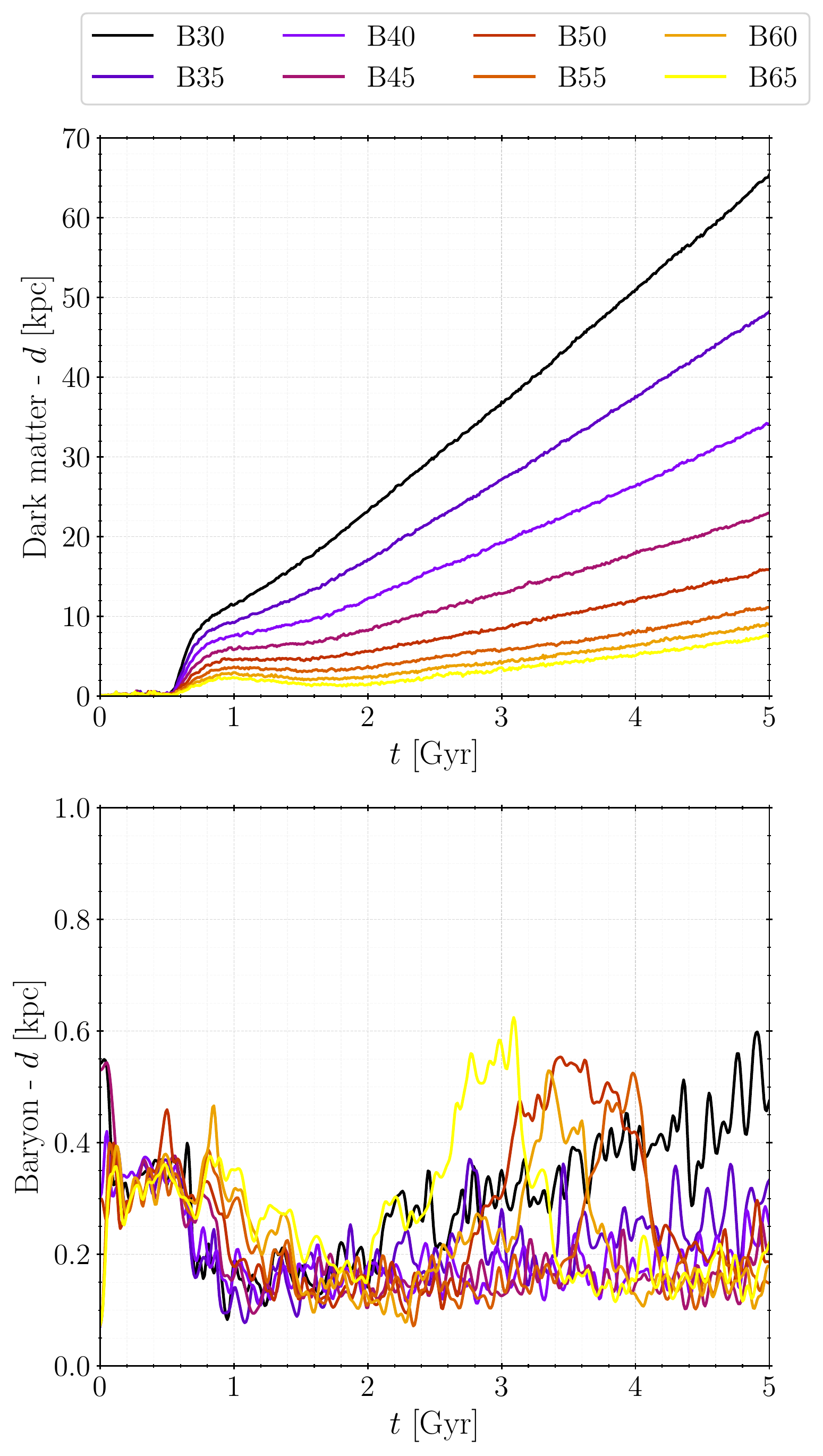}}
	\caption{Distance between two measures of intruder's center: simple center of mass and the lowest potential energy, for dark matter (upper panel) and baryon particles (lower panel).}
	\label{fig1}
\end{figure}

\indent Tidally stripped dark matter particles make the determination of the intruder's centre fairly challenging. As more dark matter particles get rejected from the intruder and shift further away from it, a simple centre of mass (which comes down to calculating the plain arithmetic mean of each coordinate for equal mass particles) also shifts away from the actual centre of the intruder. To determine the location of the actual centre, a method based on the particle's potential energy is employed. Since \texttt{GADGET2} calculates total potential energy, including contributions from the main galaxy, the densest $1$ kpc$^3$ cube with intruder particles (regardless of their type, dark matter or stellar) is first filtered. Then, the filtered particle with the lowest potential energy is chosen to represent the intruder's centre. The intruder's velocity is derived using only filtered particles.\\
\indent Figure \ref{fig1} illustrates differences between these two estimates and contributions to possible errors from each component (by calculating differences between these estimates separately for dark matter and baryon particles). Despite getting considerably lower as the impact parameter increases, these differences are still significant for dark matter halo, reaching almost $10$ kpc at a later stage. Consequently, both the estimate of dark matter halo mass and the shape of its density profile would be wrong if the classical centre of mass is used. As expected, the baryon (stellar) component appears to be unaffected and most likely retains all of its initial mass. Differences between the two centre estimates remain below $0.6$ kpc at all times in all simulations. Ideally, the centre of mass for both components should be at roughly the same position pointing to the centre of the intruder galaxy as a whole. The method used here is one way to fix issues caused by the stripped particles of dark matter halo. Alternatively, one can centre the whole intruder galaxy on the stellar centre of mass prior to any dark matter mass estimates, as stellar component does not have significant differences between the two centre estimates.\\

\subsection{Mass estimates and relevant parameters}\label{mass}

\indent After localizing the intruder's centre, three different dark matter mass estimates were calculated:
\begin{itemize}
\item \textbf{Bound mass} measure roughly estimates the gravitationally bound mass of dark matter halo. Particle velocities are centred on previously calculated intruder's velocity. Then, particles with negative total energies (potential and kinetic sum) are filtered. Despite \texttt{GADGET2} calculating total potential energy, due to high enough intruder's velocity, particles possibly captured by the primary galaxy would end up with positive total energies. However, a major pitfall of this mass measure is the inclusion of barely gravitationally bound particles forming tidal features. 
\item \textbf{Virial mass} filters dark matter particles inside intruder's virial radius. Virial radius is determined by fitting the NFW profile to the intruder's (dark matter component) density profile.
\item \textbf{Core mass} measure is based on mass estimate of \cite{klim2009,kaza2011}. First, circular velocity profile is calculated as $V_\mathrm{circ} (r) = \sqrt{G M(<r) / r}$, where $M(<r)$ is cumulative mass of dark matter halo and $r$ spherical radius. Radius $r_\mathrm{max}$ where this profile reaches maximum $V_\mathrm{max}$ is chosen as cutoff, and $M (<r_\mathrm{max})$ represents dark matter halo core mass.
\end{itemize}

\indent By definition, at later stages of simulations, virial and core mass measures should remain fairly constant. Core mass, accounting for the majority of dark matter particles, should be seen as a lower limit for total dark matter mass, whereas bound mass should represent an upper limit. Bound mass should continue to decline over time as tidal features slowly dissolve and the majority of particles become detached from the intruder. Some particles might get recaptured by the intruder, which would result in a slight increase in virial mass. Due to this, as the main measure of intruder's dark matter mass, virial mass averaged over the last 3 Gyr will be used.\\
\indent Baryon (stellar) component of the intruder is not expected to lose a significant amount of mass. Moreover, based on the previous centre of mass estimates for the stellar component, it is likely that all of its particles remain within the virial radius. However, that does not imply stellar component doesn't undergo any changes. Evolution of half-mass radius $R_{\mathrm{B},0.5}$ will be followed. Note that, with the total stellar mass remaining constant, changes in half-mass radius imply changes in the average spherical density of the stellar component: when half-mass radius increases density decreases and vice versa.\\

\section{RESULTS}\label{results}

\begin{figure*}[!ht]
\centerline{\includegraphics[width=0.99\textwidth, keepaspectratio]{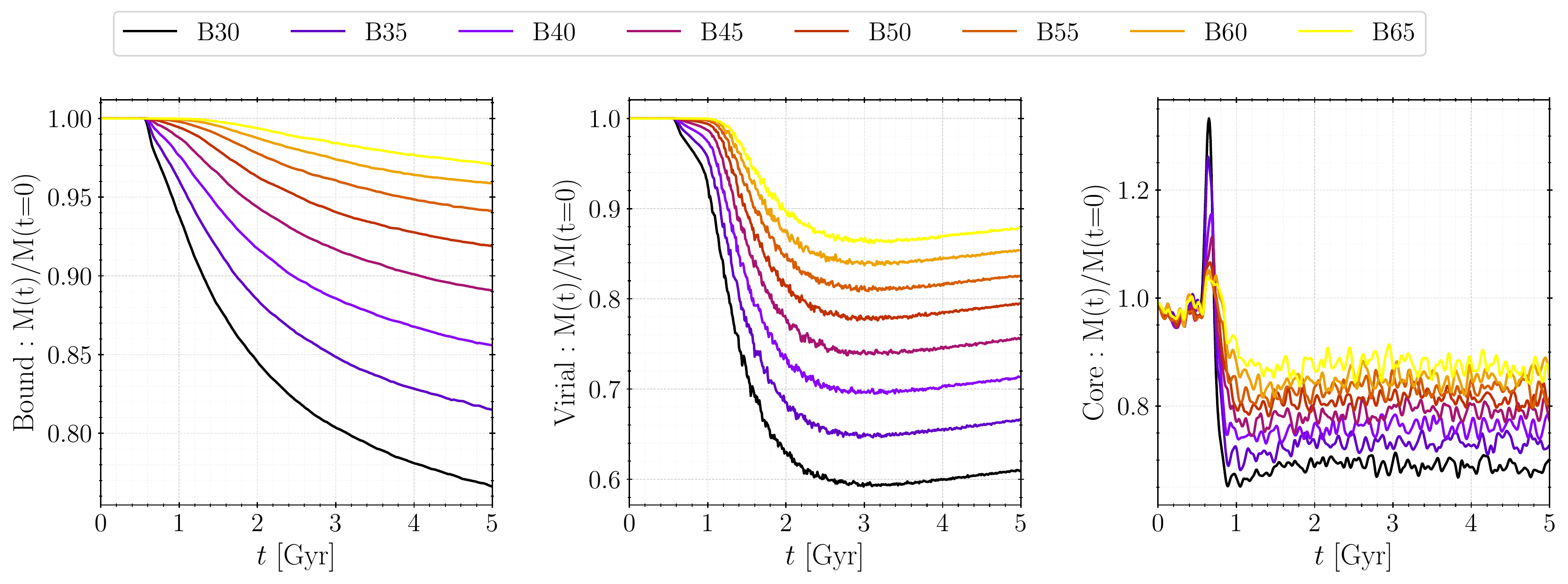}}
\caption{Evolution of dark matter halo mass estimates defined in \ref{mass} (from left to right): bound (first panel), virial (second panel), core mass (third panel). Different colors are assigned to different simulations. Estimates are expressed in relative form, compared to initial mass, and thus represent fractions of leftover mass.}
\label{fig2}
\end{figure*}

\begin{figure*}[!ht]
\centerline{\includegraphics[width=0.99\textwidth, keepaspectratio]{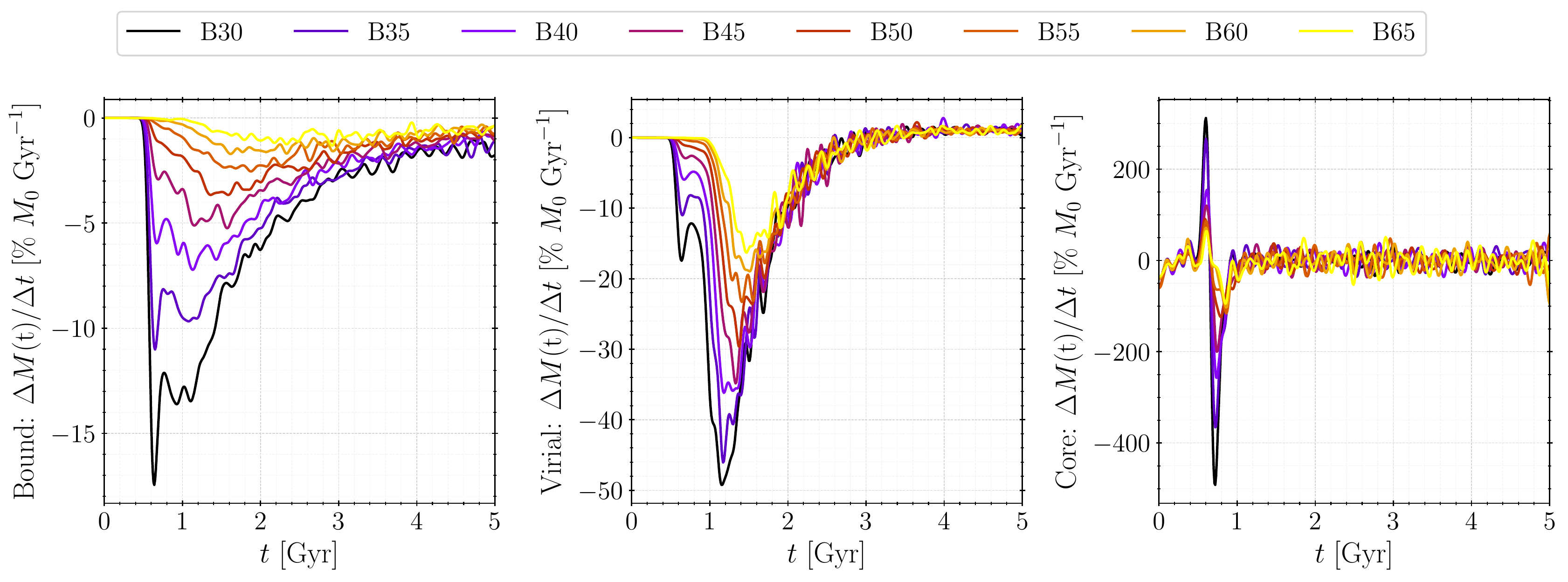}}
\caption{Mass change rate $\Delta M /\Delta t$ for different mass estimates and simulations with the same annotations as in Figure \ref{fig2}. Mass change rate is expressed as percentage of initial mass per Gyr (contrary to Figure \ref{fig2} which shows fractions).}
\label{fig3}
\end{figure*}

\indent The evolution of dark matter halo mass estimates defined in \ref{mass} is shown on Figure \ref{fig2}. These estimates are expressed in relative form as:\\
\begin{equation}
f_M = \frac{M(t)}{M(t=0)},
\label{eq:1}
\end{equation}
\noindent where $M(t)$ is the appropriate estimate (bound, virial or core) during simulation(s), $M(t=0)$ its value at the start of simulation(s), and $f_M$, thus, represents leftover fraction of initial dark matter mass. Supplementary to Figure \ref{fig2}, Figure \ref{fig3} shows the evolution of dark matter mass change rate expressed as a percentage of its initial mass per Gyr. It has to be noted that initial values of both bound and virial estimates, in all simulations, are equivalent to $M = 9.044 \cdot 10^{10} \mathrm{M}_\odot$ (initial dark matter halo total mass), while dark matter core mass equates to $M = 3.684 \cdot 10^{10} \mathrm{M}_\odot$.\\
\indent As expected, a significant mass change rate is observed after pericentre is reached for all mass estimates. During the encounter intruder stretches, becoming heavily distorted for a brief period, which is best visible on core mass plots. Core distortion is followed by abrupt mass loss, after which the dark matter core stabilizes and remains fairly constant with negligible variations until the end of every simulation. The whole process takes place before the encounter is even over. The final leftover fraction of dark matter core mass estimate is higher than the leftover fraction of virial mass estimate in most simulations (evident from Figure \ref{fig4} as well). This implies that despite the strong gravitational influence of the main galaxy, the core part of the intruder's dark matter halo remains semi-preserved. It is also the key in understanding why the baryon component can retain almost all of its initial mass - gravitational potential of preserved dark matter halo's core protects the baryon component against significant mass loss.\\
\indent Virial mass takes longer to stabilize - while dark matter mass loss starts around the time pericentre is reached, mass loss rate (Figure \ref{fig3}) peaks right after the encounter, at $t \geq 1.08$ Gyr. The peak itself shows dependence on pericentre distance, ranging from $<50\%$ of initial dark matter mass per Gyr, in simulation with the lowest pericentre distance (B30), to $\simeq 16\%$ of initial dark matter mass per Gyr, in simulation with the highest pericentre distance (B65). Following the peak, mass loss rate sharply declines from $\simeq 10\%$ of initial dark matter mass per Gyr at $t=2$ Gyr to no mass loss at $t=3$ Gyr in all simulations, irrespective of pericentre distance. After $t=3$ Gyr, there is an almost constant mass gain of $\leq 1\%$ of initial dark matter mass per Gyr in all simulations. Thus, the most significant virial mass loss is observed for $1$ Gyr following the end of the encounter. Bearing that in mind, for fitting purposes (i.e. exploring the functional relationship between leftover virial mass and impact parameter relative to the virial radius of the primary) virial mass fraction averaged over the last $3$ Gyr will be used.\\
\indent Bound dark matter mass, expectedly, declines until the end of each simulation. Its mass loss is much slower than the one of virial mass, even during peak, which happens after the pericentre is reached and before the encounter is over. During the last $1$ Gyr, mass loss rate is almost constant in all simulations, varying between $1-2\%$ of initial dark matter mass per Gyr. Given that the final fraction of bound dark matter mass is still higher than virial mass fraction at the end, and that mass loss rate is non-zero at $t=5$ Gyr, bound dark matter mass will likely continue to decline past simulation cutoff at $5$ Gyr until it converges with virial mass.\\
\indent Note that bound dark matter mass discussed here should be considered as an upper limit for truly bound mass. Other than the inclusion of particles forming tidal features outside of the virial radius, our method possibly includes dark matter particles that might not be gravitationally bound to the intruder's core. Precise determination of bound mass would, ideally, require the use of the "snowballing" method \citep{smith2015}. However, said method is robust and was not feasible, due to our limited computing resources.


\subsection{Leftover dark matter mass - dependence on impact parameter}

\begin{figure}[ht!]
	\centerline{\includegraphics[width=\columnwidth, keepaspectratio]{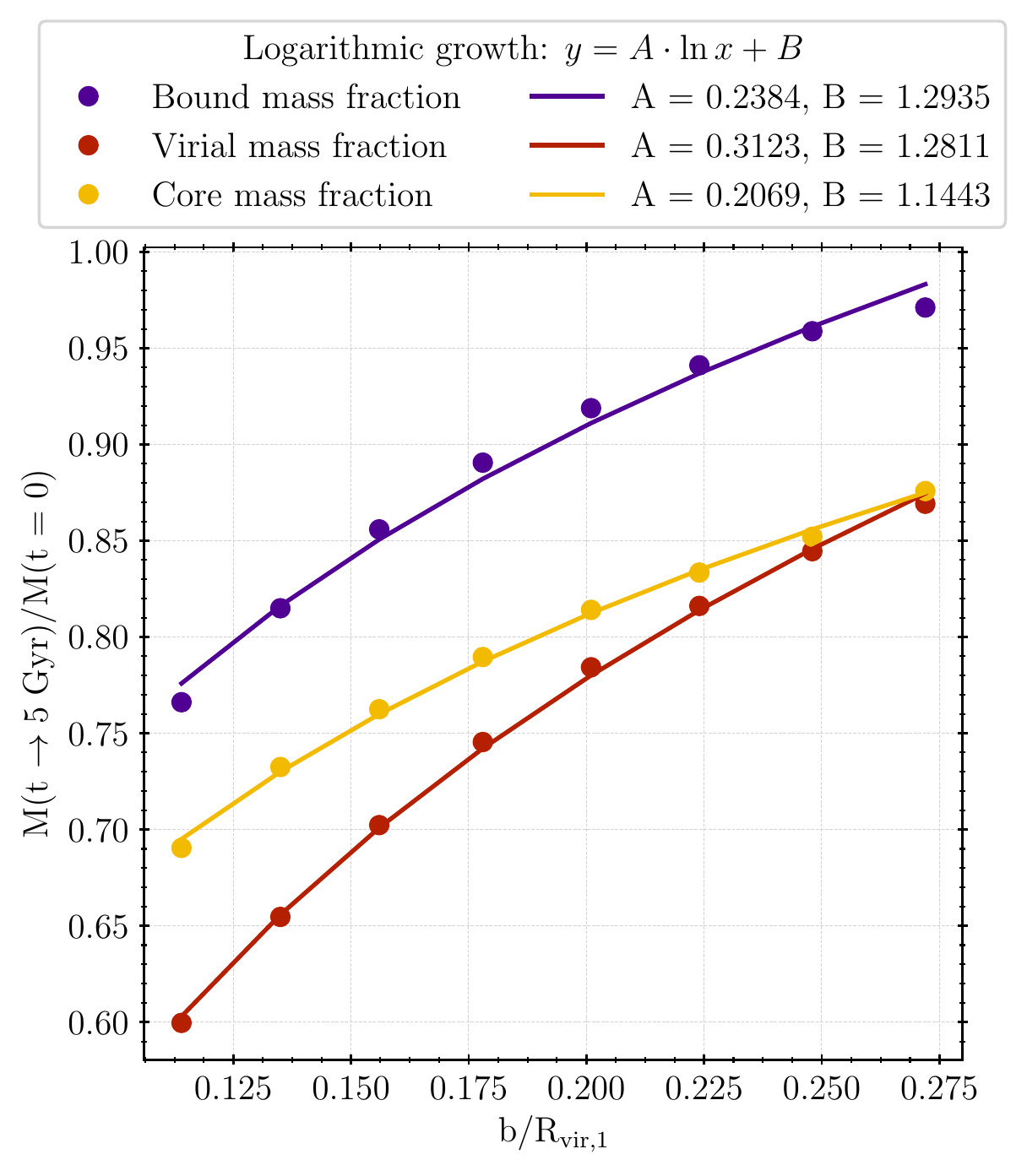}}
	\caption{Leftover mass (a fraction of initial mass) as a function of impact parameter (pericentre distance) relative to virial radius of the primary: bound (purple), virial (red), and core mass (yellow). Filled circles represent simulation data, while lines show logarithmic growth fit: $y = A \cdot \ln x + B$. Fitting parameters $A$ and $B$ are included in the legend.}
	\label{fig4}
\end{figure}

\begin{figure}[ht!]
	\centerline{\includegraphics[width=0.95\columnwidth, keepaspectratio]{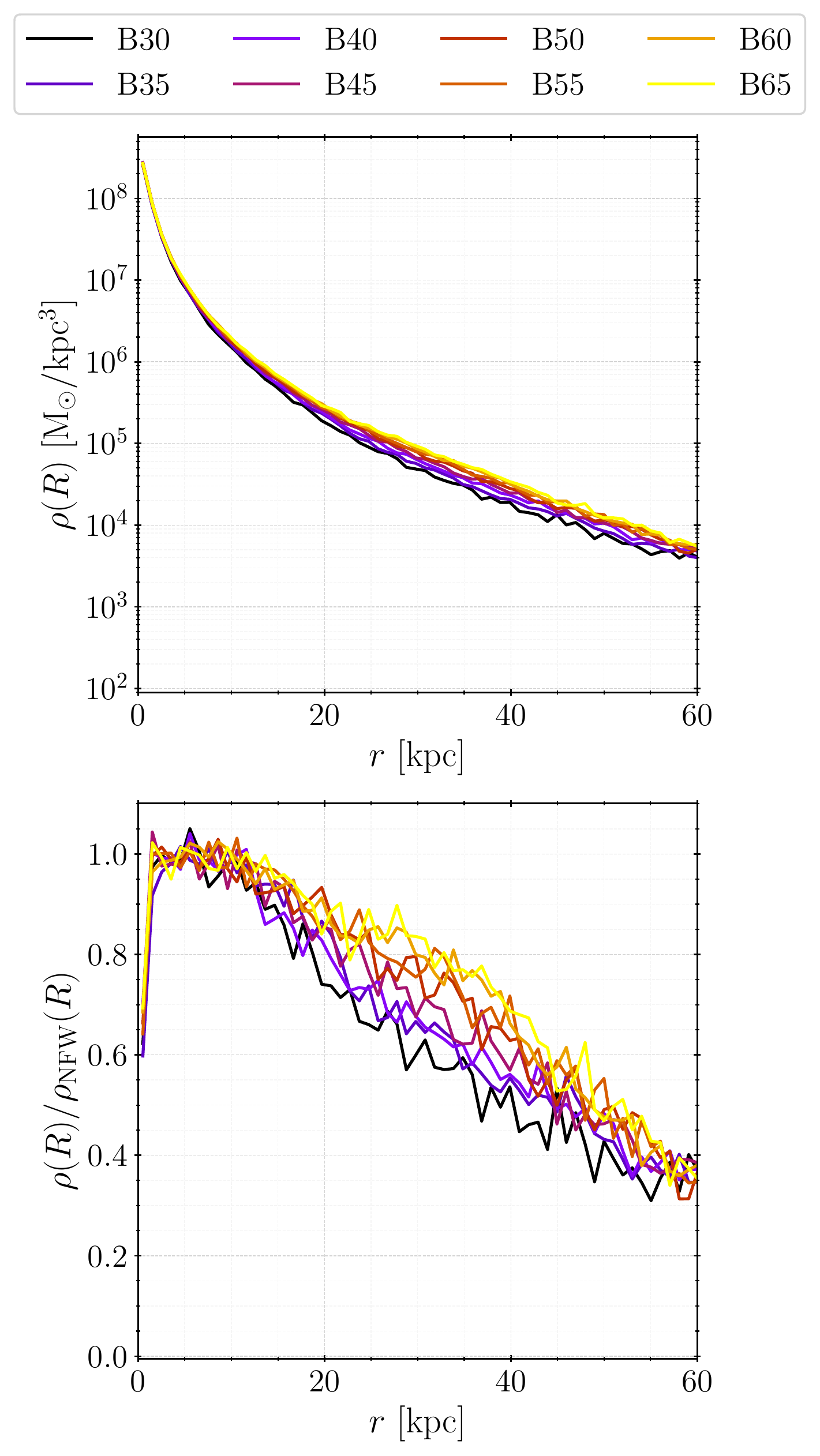}}
	\caption{\emph{Upper panel:} Final ($t = 5$ Gyr) density profiles of dark matter component. Different colors are assigned to different simulations. \emph{Lower panel:} Normalized density profiles (relative to the analytical NFW profile), with the same annotations as on upper panel.}
	\label{fig5}
\end{figure}

\indent Leftover dark matter mass fraction, for each type of mass estimate, as a function of impact parameter (pericentre distance) relative to the virial radius of the primary, is shown on Figure \ref{fig4}, with different types of mass estimates denoted by different colours. Filled circles represent simulation data, and lines represent logarithmic growth fit in a form:

\begin{equation}
    y = A \cdot \ln x + B
\end{equation}
\noindent where $y$ corresponds to the leftover dark matter mass fraction, $x$ corresponds to impact parameter (pericentre distance) relative to the virial radius of the primary, and $A$ and $B$ are fitting parameters. As evident, virial and core leftover dark matter mass's dependence on impact parameter is perfectly described with logarithmic growth law. Bound dark matter mass fraction, albeit deviating from the fitted line, can still be described with logarithmic growth law. Fitting parameters are different for all three types of mass: virial mass has $A=0.3123$ and $B=1.2811$, core mass has $A=0.2069$ and $B=1.1443$, and bound mass has $A=0.2384$ and $B=1.2935$. These fitting parameters are likely not universal but dependent on multiple interaction parameters, e.g. the mass ratio of interacting galaxies, the initial relative velocity of intruder galaxy, interaction duration.\\
\indent Surprisingly, outside-in nature of the tidal stripping, which is one of the main formation mechanisms of ultra-compact dwarf galaxies (mentioned in the introductory part of this paper), is less evident in flybys with larger impact parameters (i.e. weaker ones). Examining extreme cases, simulation B30 with the lowest and simulation B65 with the highest impact parameter yields interesting insights. In B30, the leftover dark matter core mass fraction is by $\sim 9\%$ higher than the leftover virial mass fraction, while they are almost equal in B65. Generally, decreasing difference between these two dark matter mass estimates with increasing impact parameter indicates that outer parts of dark matter halo are more affected in closer (i.e. stronger) flybys and thus lose more mass than inner (core) parts. As the impact parameter increases, dark matter mass loss becomes almost uniform with radius. This might lead to an assumption that density profiles, and thus shapes and slopes of NFW profiles, are heavily affected. In reality, this is not entirely the case (Figure \ref{fig5}) - while differences become visible on larger radii ($R \geq 18$ kpc), density profiles are not significantly altered in inner parts where the majority of leftover particles reside. Normalized density profiles (relative to their fitted analytical NFW profile) are shown on the lower panel on Figure \ref{fig5}. High deviations at the centre are understandable as the analytical form has unrealistically high densities on low radii ($R\rightarrow 0$). Well outside dark matter half-mass radius ($R \geq 18$ kpc)\footnote{Half-mass radius, which encloses half of the leftover dark matter mass, varies between $\sim 5.5$ kpc (B30) and $\sim 7.5$ kpc (B65). These can be approximated for each simulation using the data listed in Table \ref{tab3} - namely leftover mass $M_\mathrm{DM}$, virial radius $R_\mathrm{vir,2}$ and concentration parameter $c$.}, deviations of density profiles from the analytical NFW increase drastically. The effect is more prominent in closer flybys where higher fractions of dark matter mass are stripped. This confirms that, due to the outside-in nature of the tidal stripping, outer parts of the dark matter halo density profiles are steeper than the analytical NFW \citep[e.g.][]{Okamoto1999, Genina2021}. Thus, the NFW does not describe the dark matter halo density profiles well overall, while the approximation is still consistent in the inner parts.\\
\indent Dependence of leftover dark matter mass (of intruder galaxy in flybys) on impact parameter is, undisputedly, perfectly described with logarithmic growth law. Naturally, the lost mass would then follow the exponential decay law. As such, decreasing impact parameter is followed by considerable and ever so faster dark matter mass loss. This will be discussed in detail, while keeping in mind constraints of this work, in Section \ref{conclusion}

\subsection{Changes to stellar component and dark-to-stellar mass ratio}

\begin{figure}[ht!]
	\centerline{\includegraphics[width=0.96\columnwidth, keepaspectratio]{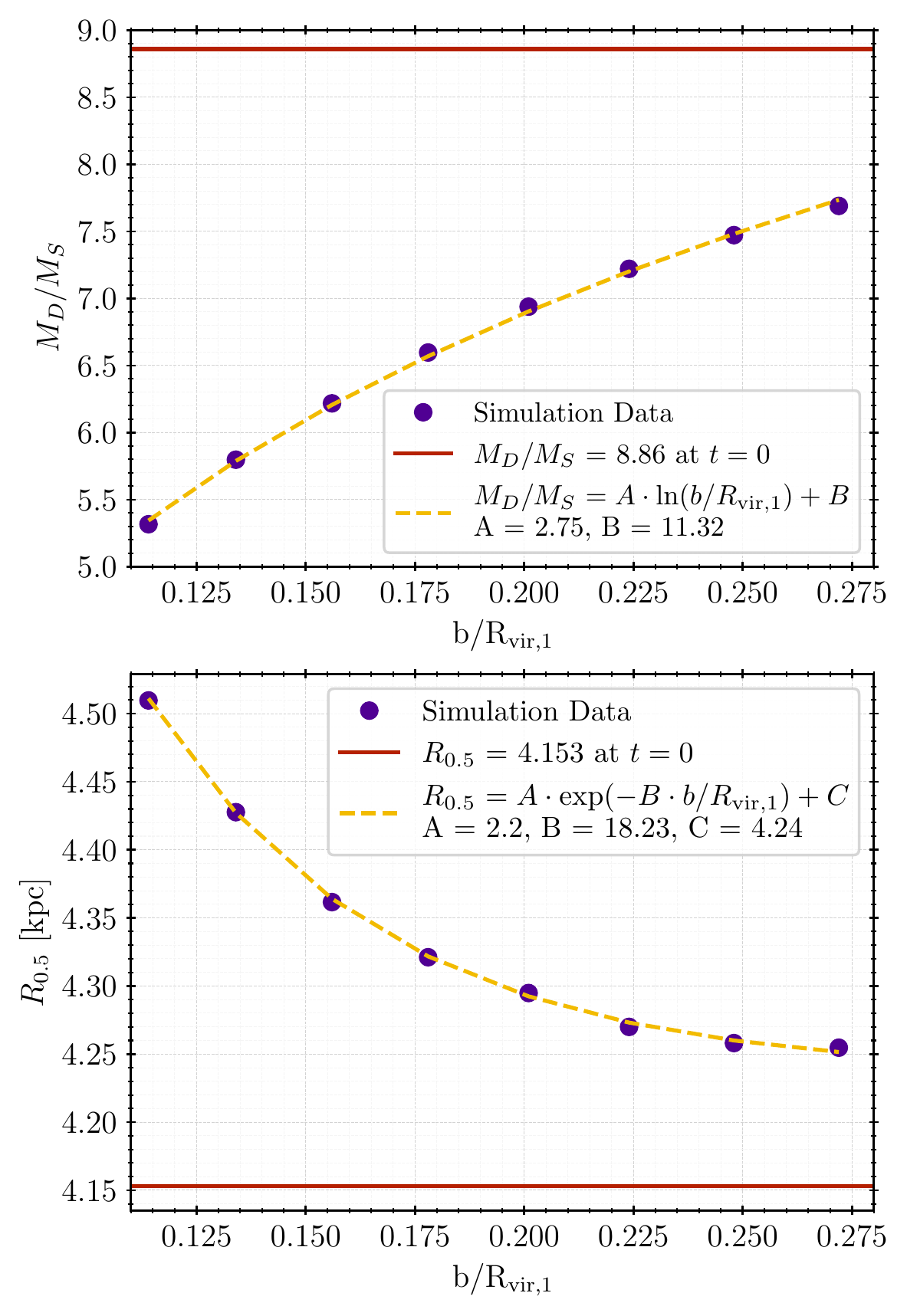}}
	\caption{On both panels, filled circles represent simulation data, solid line initial value (at $t = 0$), and dashed line best fit curve, with fitting functions and parameters included in the legend; as a function of impact parameter relative to virial radius of the primary. \emph{Upper panel}: Dark-to-stellar mass ratio, where $M_{D}$ is final (at $t = 5$ Gyr) virial mass, and $M_{S}$ final mass of the stellar component. \emph{Lower panel}: Final half-mass radius of the stellar component, $R_{0.5}$.}
	\label{fig6}
\end{figure}

\indent Contrary to the dark matter halo, the stellar component does not suffer any mass loss. The total stellar mass was estimated by applying the method for bound dark matter mass estimate to stellar particles. Since it remains constant in all simulations, visualisation of it is omitted. However, the total stellar mass remaining constant does not imply there are no changes to the stellar component. Final half-mass radius of the stellar component, $R_{0.5}$, as a function of impact parameter (relative to the virial radius of the primary), is shown on the lower panel of Figure \ref{fig6}. Exponential decay law, $R_{0.5} = A \cdot \mathrm{exp}(-B\cdot b/R_\mathrm{vir,1}) + C$ describes its behaviour the best, with fitting parameters: $A = 2.2$, $B = 18.23$, and $C = 4.24$. While it stretches faster as impact parameter decreases, the value of $R_{0.5} = 4.5097$ kpc in the strongest flyby simulation (B30) is not considerably larger than initial one of $R_{0.5} = 4.153$ kpc. Typical flybys, hence, can contribute to the formation of ultra-diffuse galaxies but cannot be its sole formation mechanism.\\
\indent Unsurprisingly, given that stellar component does not suffer mass loss, dark-to-stellar mass ratio $M_{D}/M_{S}$ shown on the upper panel of Figure \ref{fig6} follows logarithmic growth law (similar to that of leftover virial mass), with fitting parameters $A = 2.75$ and $B = 11.32$. From initial value of $M_{D}/M_{S} = 8.86$, this ratio drops to $M_{D}/M_{S} = 5.317$ in the most extreme case (simulation B30). Despite losing almost half of its initial dark matter mass, the intruder galaxy remains dark matter dominated, albeit less so. Variations in this ratio caused by galaxy flybys, however, can contribute to the scatter in SHMR (stellar-to-halo mass relation) at the lower mass end, $M_\mathrm{halo} \leq 10^{11} M_{\odot}$. Scatter in SHMR in this, dwarf regime has received very little attention so far. Its importance is becoming more clear, as extreme cases could successfully explain the formation of dark matter deficient galaxies \citep{trujillogomez2021}.\\

\begin{table*}
	\caption{{List of the most relevant results, where $b$ is pericentre distance, $R_{\mathrm{vir},1}$ virial radius of main galaxy, $M_{\mathrm{DM}}$} virial leftover mass expressed as a percentage of initial virial mass, $M_{\mathrm{DM}}/M_{\mathrm{B}}$ dark-to-baryon mass ratio, $R_{\mathrm{vir},2}$ virial radius of intruder galaxy averaged over last 3 Gyr, $\Delta R_{\mathrm{vir},2}$ variation of intruder's virial radius over last 3 Gyr, $R_{\mathrm{B},0.5}$ half-mass radius of baryon (stellar) component at the end of simulation, and $c$ dark matter halo concentration parameter at $t=5$ Gyr.}
	\label{tab3}
	\vskip.25cm
	\centerline{\begin{tabular}{ccccccccc}
			\hline
			Name & $b$ [kpc] & $b/R_{\mathrm{vir},1}$ & $M_{\mathrm{DM}}$ [\%] & $M_{\mathrm{DM}}/M_{\mathrm{B}}$ & $R_{\mathrm{vir},2}$ [kpc] & $\Delta R_{\mathrm{vir},2}$ [kpc] & $R_{\mathrm{B},0.5}$ [kpc] & c\\
			\hline
			B30 &	22.5	& 0.114 &	59.96	& 5.317	& 87.95	& 0.92	& 4.5097 & 28.94\\
			B35	& 26.53	& 0.135	& 65.46	& 5.798	& 89.74	& 1.03	& 4.4277 & 29.35\\
			B40	& 30.69	& 0.156	& 70.23	& 6.218	& 92.48	& 1.09	& 4.3616 & 26.62\\
			B45	& 35.07	& 0.178	& 74.55	& 6.597	& 93.69	& 1.17	& 4.3211 & 26.47\\
			B50	& 39.62	& 0.201	& 78.43	& 6.939	& 94.75	& 1.04	& 4.2948 & 26.56\\
			B55	& 44.27	& 0.224	& 81.61	& 7.221	& 95.83	& 1.14	& 4.2700 & 26.10\\
			B60	& 48.99	& 0.248	& 84.45	& 7.472	& 96.90	& 1.10  & 4.2581 & 25.63\\
			B65	& 53.72	& 0.272	& 86.92	& 7.690	& 97.77	& 1.09	& 4.2547 & 24.94\\
			\hline
	\end{tabular}}
\end{table*}

\section{DISCUSSION AND CONCLUSION}\label{conclusion}

\indent Series of N-body simulations of typical (10:1 mass ratio), deep galaxy flybys was performed, with differing impact parameters, ranging from $0.114$ to $0.272$ of the virial radius of the primary galaxy. The focus of the analysis was on dark matter mass loss of the secondary, intruder galaxy with three different methods of estimating the total dark matter mass of the intruder. Dependence on impact parameter for all of those, for leftover dark matter mass, is that of logarithmic growth, and exponential decay for lost dark matter mass. This functional dependence seems universal, while fitting parameters may vary with different initial conditions and interaction parameters (such as the mass ratio of interacting galaxies, the initial relative velocity of intruder galaxy and thus interaction duration). Most relevant visualised results in this paper are listed in Table \ref{tab3} as a summary.\\
\indent Bound dark matter mass calculated here should be taken as an absolute upper limit of truly bound dark matter leftover mass, given the rough nature of this estimate. Furthermore, such an estimate has very little applicability. Virial mass, conversely, is a much better indicator of intruder galaxy's total dark matter mass as it's suitable for comparison with results and data of cosmological simulations. However, this value should be taken with a dose of scepticism due to considerable deviations of dark matter density profiles from the analytical NFW on higher radii. Still, it is not the only noteworthy dark matter mass estimate. Core dark matter mass, estimating the total mass of the inner dark matter halo, is an appropriate one for comparison with observationally derived dark matter fractions, which can typically only probe regions where baryons are present. Particularly of interest is that this estimate stabilizes faster than the virial one, reaching its final values even before the encounter is over. This makes it convenient to estimate, to an extent, total virial mass and its loss which happens $1$-$2$ Gyrs later, based on observationally derived dark matter masses. Disparities may, of course, arise due to the outside-in nature of tidal stripping, especially in closer flybys.\\
\indent Typical flybys investigated here could not be the sole culprit behind the formation of ultra-diffuse or dark matter deficient galaxies, but their effects and contributions should not be disregarded. Considering their frequency at the present epoch \citep{sinha2012,shan2019}, combined with other possible close encounters and collective effects of galaxy clusters, where such events are likely to take place, these scenarios become highly plausible.\\
\indent Given the nature of exponential decay with impact parameter, of both lost dark matter mass, and half-mass radius of the stellar component, it is fairly safe to assume that closer (or stronger in a different way, e.g. slower) flybys than the ones investigated here, could alone lead to the formation of ultra-diffuse and dark matter deficient galaxies with \emph{just the right} impact parameters. The formation of ultra-diffuse galaxies might not require a narrow range of impact parameters. Stellar component stretches at a faster rate than the dark matter mass loss. While stretching and getting extended, it might become prone to the mass loss itself, although that usually only starts happening after a significant portion of dark matter ($\sim 80\%$) is already stripped \citep{smith2016,lokas2020}. Since this was not observed even in the simulation with the closest flyby presented here, an estimate of the required impact parameter for such a case would be pure speculation at this point.\\
\indent The possible formation of dark matter deficient galaxies through galaxy flybys is much more sensitive. It would require flybys with extremely low impact parameters, or much slower deep flybys, in addition to the faster rate of dark matter stripping compared to stellar one. Moreover, it might require specific shapes of density profiles for both dark matter, and baryon component. While such flybys are extremely rare, they are still detected in cosmological simulations \citep{sinha2015}, which is in line with the exotic nature of these objects.\\
\indent This is, of course, entirely speculative. The approach and methods, presented in this work, are unfit to deal with flybys with lower impact parameters. Simulations being pure N-body are rather simplified, and have no way of dealing with various and complex physical processes which would occur in such interactions. Rare, atypical and stronger (in any way, with lower impact parameters, lower initial velocities or higher differences in initial masses of interacting galaxies) flybys are, however, worth exploring further. The best approach would be trying to reconcile complex hydrodynamical simulations of isolated flybys with cosmological simulations, and hopefully observational data.\\


\acknowledgements{I am grateful to my mentor, Miroslav Mi{\' c}i{\' c}, for his immense support and patience. I would also like to thank the reviewer for the insightful comments that helped improve the quality of this work. This work was supported by the Ministry of Education, Science and Technological Development of the Republic of Serbia (MESTDRS) through the contract no. 451-03-9/2021-14/200002 made with Astronomical Observatory of Belgrade, and the contract no. 451-03-9/2021-14/200104 made with Faculty of Mathematics, University of Belgrade. The python packages \texttt{MATPLOTLIB} \citep{Hunter2007}, \texttt{NUMPY} \citep{Harris2020}, \texttt{SCIPY} \citep{Virtanen2020}, \texttt{PANDAS} \citep{McKinney2010}, and \texttt{PYNBODY} \citep{pynbody} were all used in parts of this analysis.}


\vskip2mm

\newcommand\eprint{in press }

\bibsep=0pt

\bibliographystyle{aa_url_saj}

{\small

\bibliography{am_paper}
}



\clearpage

{\ }


{\ }

\newpage

\begin{strip}
	
{\ }



\naslov{GUBITAK MASE TAMNE MATERIJE U PROLETIMA GALAKSIJA: ZAVISNOST OD PARAMETRA SUDARA}


\authors{A Mitra{\v s}inovi{\' c}$^{1,2}$}

\vskip3mm


\address{$^1$Astronomical Observatory, Volgina 7, 11060 Belgrade 38, Serbia}

\address{$^2$Department of Astronomy, Faculty of Mathematics,
University of Belgrade\break Studentski trg 16, 11000 Belgrade,
Serbia}


\Email{amitrasinovic@aob.rs}

\vskip3mm


\centerline{{\rrm UDK} \udc}


\vskip1mm

\centerline{\rit Originalni nauqni rad}

\vskip.7cm




\begin{multicols}{2}

{
\rrm

Proleti galaksija, interakcije gde dva nezavisna haloa medjusobno prodiru, ali se zatim odvoje i ne sudare, su qesta pojava na ni{\zz}im crvenim pomacima. Ove interakcije mogu imati znaqajan uticaj na evoluciju pojedinaqnih galaksija, poqev od gubitka mase i promena oblika galaksija, pa do pojave plimskih efekata i svojstava, i formiranja razliqitih morfolo{\ss}kih struktura u diskovima galaksija. Glavni fokus ovog rada je na gubitku mase tamne materije sekundarne, "uljez" galaksije, sa ciljem odredjivanja funkcionalne zavisnosti gubitka mase tamne materije od parametra sudara. Serija simulacija N-tela tipiqnih proleta galaksija (odnos masa 10:1) sa razliqitim parametrima sudara, pokazuje da preostala masa tamnog haloa "uljez" galaksije prati zakon logaritamskog rasta sa parametrom sudara, bez obzira na naqin na koji je ukupna masa procenjena. Izgubljena masa onda, jasno, prati zakon eksponencijalnog raspada. Zvezdana komponenta se br{\zz}e {\ss}iri sa smanjenjem parametra sudara, prate{\cc}i zakon eksponencijalnog raspada. Funkcionalna zavisnost od parametra sudara u svim sluqajevima se qini univerzalna, ali su njeni odgovaraju{\cc}i parametri verovatno osetljivi na parametre interakcije i poqetne uslove (npr. odnos masa interaguju{\cc}ih galaksija, relativna poqetna brzina "uljez" galaksije, trajanje interakcije). Dok tipiqni proleti, prouqavani ovde, ne mogu biti jedini uzroqnik formiranja ultra-difuznih galaksija, ili galaksija bez velike koliqine tamne materije, njihovi efekti ne smeju biti ignorisani jer mogu bar znaqajno da doprinesu ovim pojavama. Stoga su retki, netipiqni, i jaqi proleti vredni daljeg i detaljnijeg istra{\zz}ivanja.

{\ }

}

\end{multicols}

\end{strip}


\end{document}